# "Strange attractor like" electric seismic precursor observed prior to the Nafpactos, Greece, EQs of 18th / 22nd January, 2010 (Ms = 5.7R / 5.6R).

Thanassoulas[1], C., Klentos[2], V., Verveniotis, G[3].

1. Retired from the Institute for Geology and Mineral Exploration (IGME), Geophysical Department, Athens, Greece.
   e-mail: thandin@otenet.gr - URL: www.earthquakeprediction.gr

2. Athens Water Supply & Sewerage Company (EYDAP),
   e-mail: klenvas@mycosmos.gr - URL: www.earthquakeprediction.gr

3. Ass. Director, Physics Teacher at 2nd Senior High School of Pyrgos, Greece.
   e-mail: gver36@otenet.gr - URL: www.earthquakeprediction.gr

**Abstract.**

Phase maps, of the earth's electric field monitored at PYR and ATH monitoring sites in Greece, were compiled for two distinct oscillating earth's electric field components long before, during and after the Nafpactos EQs in Greece (January 18th and 22nd of 2010, Ms = 5.7 / 5.6R). The selected periods of the earth's oscillating electric field were the T = 1 day and T = 14 days. The specific components were selected due to the fact that they are highly depended from the corresponding same period tidal waves (M1, K1, P1). It was found that the "strange attractor like" electric seismic precursor preceded the main seismic event for four days for the case of T = 1 day, while a much longer period of time (11 days) was observed for the case of T = 14 days. It seems that the shorter the used wavelength is the shorter is the predictive time window regarding the pending seismic event. The present work replicated earlier similar results obtained by the same methodology for past large seismic events in Greece and have already been presented in the same database. The present methodology seems to be a promising tool in short-term (in time) earthquake prediction.

## 1. Introduction.

Generally, the term "earthquake prediction" refers to the determination of the three prognostic parameters of an earthquake at some time before its occurrence. These parameters are: the time of occurrence, the epicenter location and the magnitude. It is obvious that for an actual prediction what is more required is the error window of each determined parameter. From a geological – tectonic and seismological point of view, there is more or less knowledge about the seismogenic areas which have already in the past generated large earthquakes. Consequently, assuming a specific dangerous seismogenic area and moreover considering only large earthquakes, what is left for determination is the time of occurrence of the next to come future large earthquake.

Furthermore, depending on the determined time window length in which the future large earthquake is expected to take place, the "earthquake prediction" is defined as: long term, medium term and short term according to time windows of some decades of years, some years and of anything less than a year. Usually, the first two categories are treated by statistical methods. Related references on these methods can be found in the monograph "Short-term earthquake prediction" (Thanassoulas, 2007).

Quite new methods in time prediction were introduced recently, by making use of the seismic electric signals. The VAN team combined the time of generation of seismic electric signals (SES) with statistical methods applied on the subsequent earthquakes generated within a specified seismogenic area which, in turn, was deduced by the observed SES signals (Varotsos 2005, Sarlis et al. 2008) aiming into shortening the occurrence time window of a future large EQ.

A different approach was followed by other researchers. It was shown that the intersection of the intensity vectors of the oscillating earth's electric field, determined at two different monitoring sites, behaved quite differently long before the occurrence of a large earthquake, short (a few days) before its occurrence and after its occurrence. In detail, the intensity vectors intersections, as a function of time at each monitoring site, formed random hyperbolas in planar space when the seismogenic area was well before its last stress load phase before the earthquake occurrence and after its occurrence, while it formed ellipses when the seismogenic area was very short (a few days) before the earthquake occurrence (Thanassoulas 2007, Thanassoulas et al. 2008a,b, 2009a,b,c). Moreover, this peculiar behavior of the earth's electric field was identified in different oscillating components of it. Examples for periods of T = 1 day, T = 14 days and T = 6 months were presented by Thanassoulas et al. (2008b, 2009a, b).

In this work we study the behavior of the earth's electric field monitored by the ATH and PYR monitoring sites (fig. 1) as follows. The study is performed on the oscillating earth's electric field for the following cases of: T = 1 day (time period from January 1st 2010 to February 9th 2010) and T = 14 days (time period from December 10th 2009 to February 9th 2010). Actually we investigate the formation of hyperbolas or ellipses in this time period and compare them to the time of occurrence of the two Nafpactos EQs (18th / 22nd January, 2010, Ms = 5.7R / 5.6R).

## 2. Data presentation and analysis.

Four earthquakes occurred with a magnitude of Ms ≥ 5.0R in the entire study period, (December 10th 2009 to February 9th 2010), which are tabulated in the following Table-1. Date is given in yyyymmddhhmm form.

### TABLE – 1

| No. | Date | Lat. | Lon. | Z | Ms |
|---|---|---|---|---|---|
| 1 | 201001140425 | 39.24 | 22.27 | 88 | 5.0 |
| 2 | 201001172016 | 35.26 | 27.86 | 29 | 5.5 |
| 3 | 201001181556 | 38.41 | 21.95 | 20 | 5.7 |
| 4 | 201001220046 | 38.42 | 21.97 | 12 | 5.6 |

The four EQs which occurred are numbered from 1 to 4 and their location is shown in the following map of Greece of figure (1).

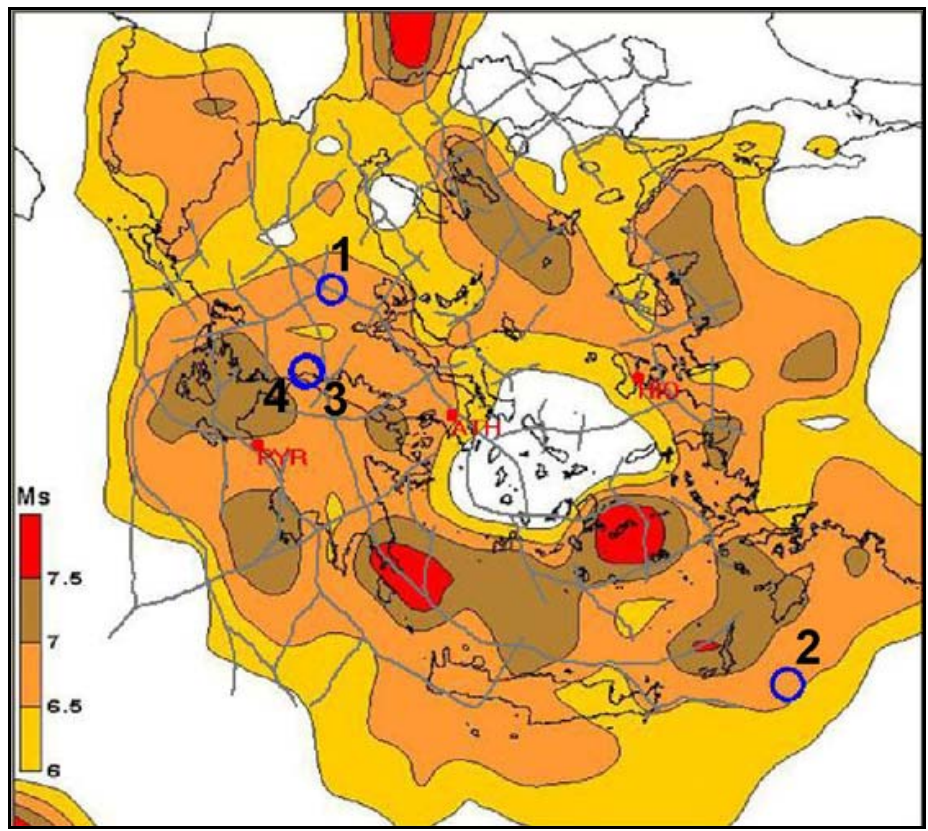

Fig. 1. Location (blue circles) of the EQs (1 to 4) with Ms ≥ 5.0R in Greece. Colored areas indicate the seismic potential spatial distribution in terms of expected EQ magnitude according to the attached color scale. The thick gray lines indicate the deep lithospheric fracture zones and faults. PYR and ATH monitoring sites are denoted by red capital letters.

The depth of occurrence of the referred EQs ranges from 88km (no.1) to 12km (no.4). The latter plays an important role concerning the generation of preseismic electric signals (Thanassoulas, 2007).
The earth's electric field data window which will be analyzed in this work is selected in such a way so that it covers, in both cases (T=1 day, T=14 days), a quite large time period before and after the occurrence of the related EQs. Therefore, the comparison of the preseismic behavior (close to the seismic event) of the earth's electric field will be more easily utilized to the time periods when there is no lithospheric strain charge increase preparatory phase for any pending future large seismic event.
The earth's electric field oscillating components will be studied separately as follows:

### 2.1. Oscillating electric field of T = 1 day

The normalized, to E-W and N-S directions, earth's electric field which was recorded from PYR and ATH monitoring sites was band-pass (T=1 day) filtered and used to generate the corresponding phase maps (Thanassoulas et al. 2008a) for the time period from January 1st to February 9th 2010. In the following figures (2, 3, 4, 5) are presented the normalized raw data for each monitoring site and the band-pass filtered oscillating component, for T = 1 day, earth's electric field. In all these figures, the occurred EQs of Ms ≥ 5.0R are denoted with red bars.

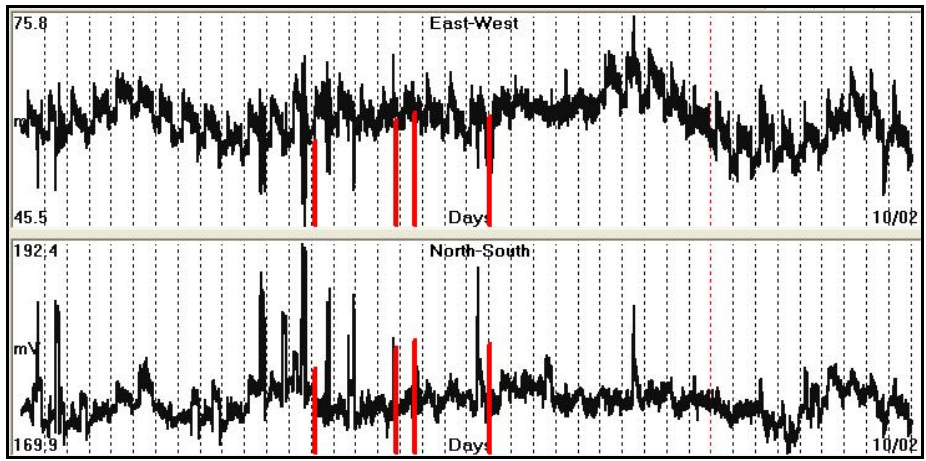

**Fig. 2. ATH normalized raw data. Black line = normalized earth's electric field components. Red bars = occurred EQs of Ms ≥ 5.0R. Data spans from: January 1st 2010 to February 9th 2010.**

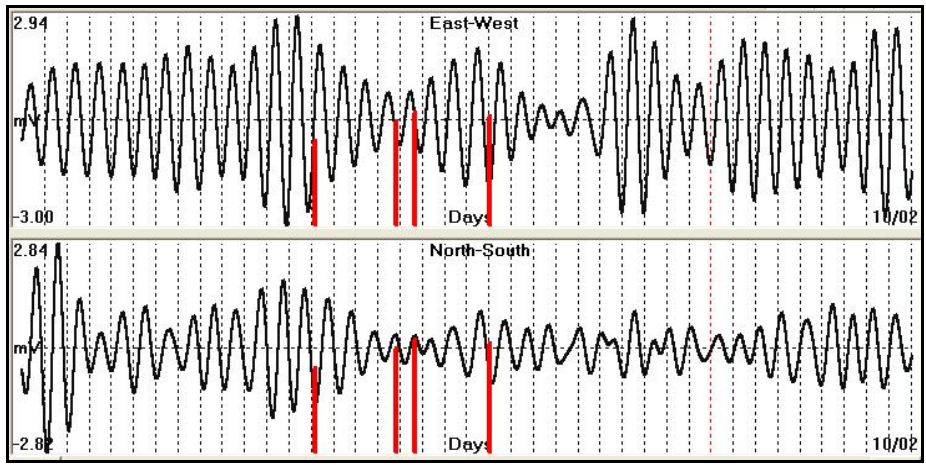

**Fig. 3. ATH normalized oscillating data. Black line = normalized earth's electric field components. Red bars = occurred EQs of Ms ≥ 5.0R. Data spans from: January 1st 2010 to February 9th 2010.**

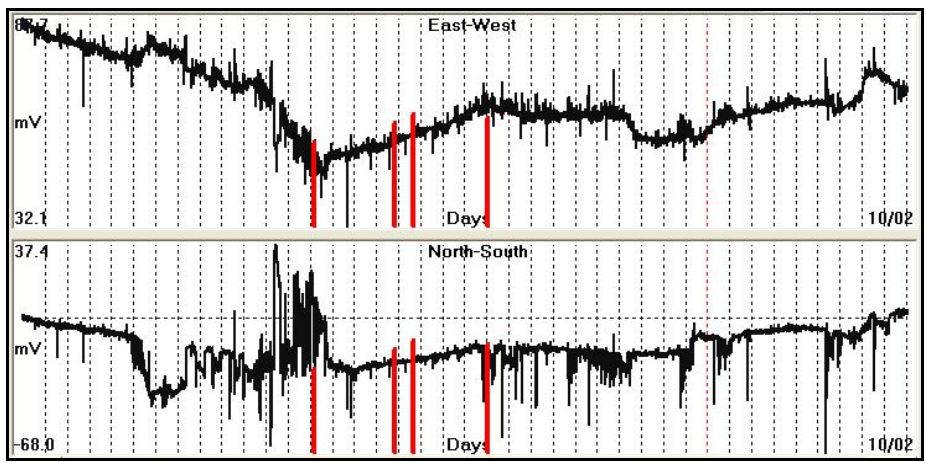

**Fig. 4. PYR normalized raw data. Black line = normalized earth's electric field components. Red bars = occurred EQs of Ms ≥ 5.0R. Data spans from: January 1st 2010 to February 9th 2010.**

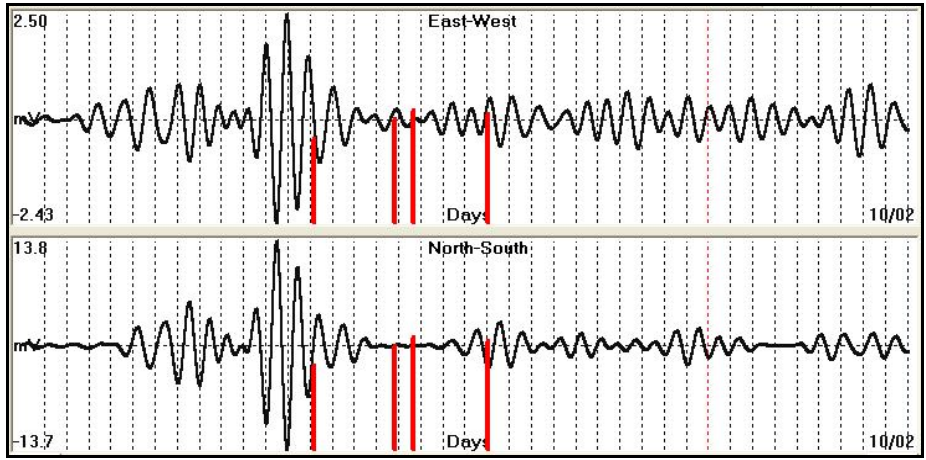

**Fig. 5. PYR normalized oscillating data. Black line = normalized earth's electric field components. Red bars = occurred EQs of Ms ≥ 5.0R. Data spans from: January 1st 2010 to February 9th 2010.**

**The data presented in figures (3) and (5) were used to compile the following phase maps presented in figures (6, 7, 8, 9).**

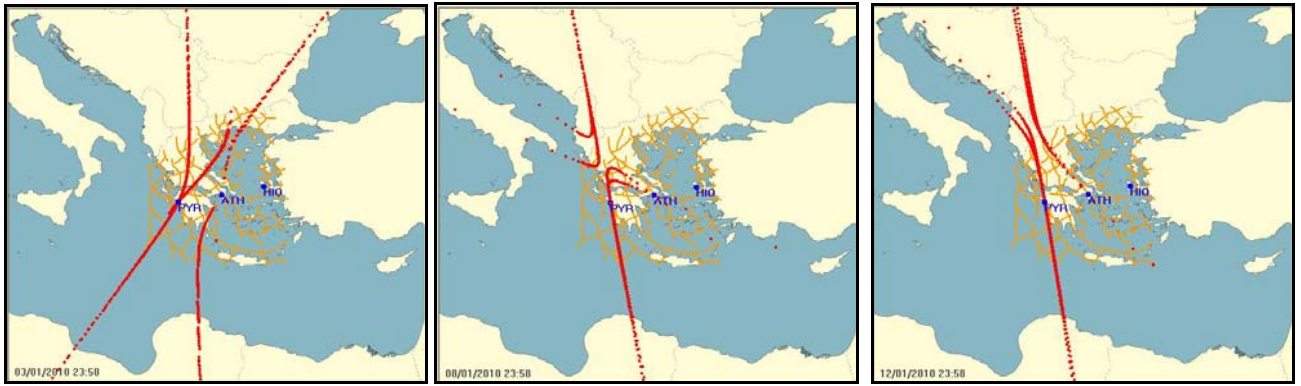

**Fig. 6. Phase maps (hyperbolas) compiled for January 3rd (left), January 8th (middle) and January 12th (right) of 2010.**

Figure (6) shows the kind of phase maps (random hyperbolas) which are generated when the seismogenic area of the future large EQ is still far from being critically strain charged and thus far (in time) from the occurrence time of the expected large EQ.

Figures (7) and (8) present the typical phase maps which are generated when the candidate to be seismically activated seismogenic area has reached critical strain charge conditions very short (a few days) before the EQ occurrence time.

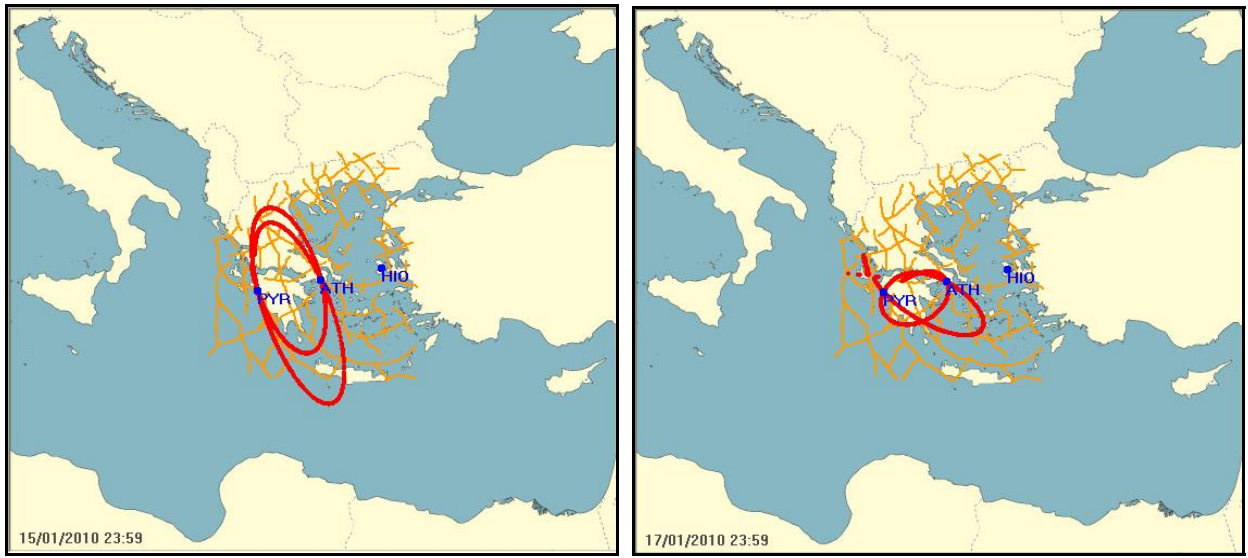

**Fig. 7. Phase maps (ellipses) compiled for January 15th (left) and January 17th (right) of 2010.**

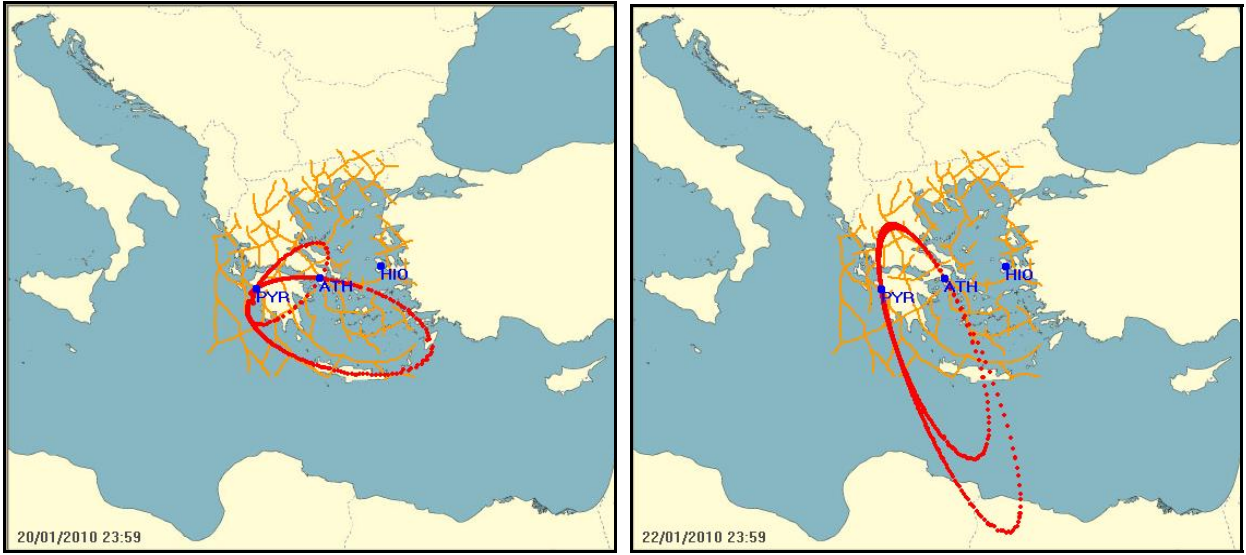

**Fig. 8.** Phase maps (ellipses) compiled for January 20$^{th}$ (left) and January 22$^{nd}$ (right) of 2010.

It is worth to notice that the EQs presented in Table – 1 did occur during the time period when the phase maps were characterized by ellipses.

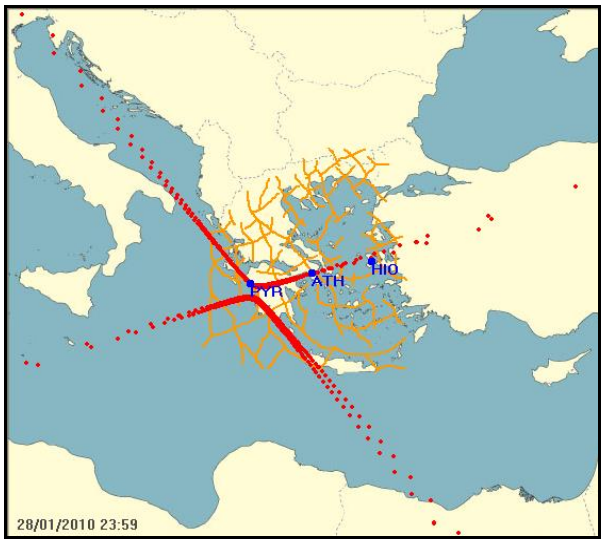

**Fig. 9a.** Phase maps (hyperbolas) compiled for January 28$^{th}$ of 2010.

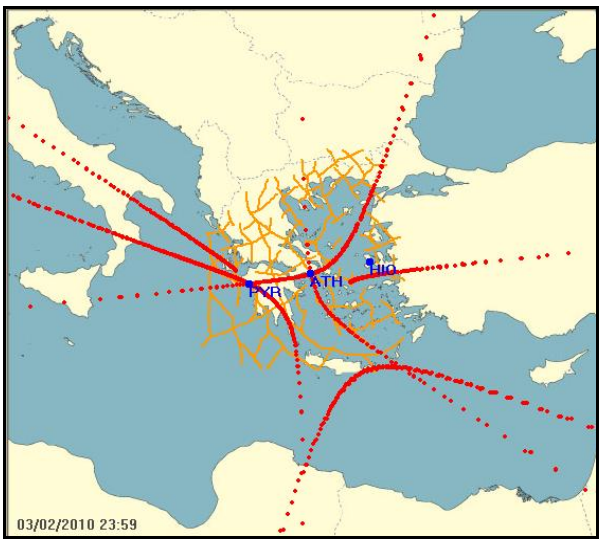

**Fig. 9b.** Phase maps (hyperbolas) compiled for February 3$^{rd}$ of 2010.

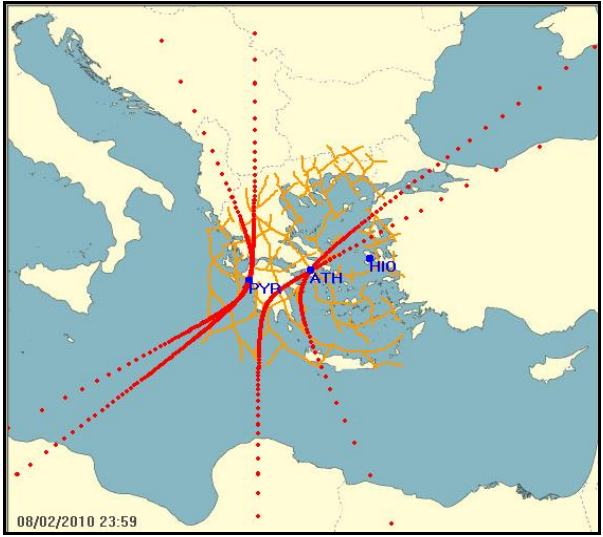

**Fig. 9c. Phase maps (hyperbolas) compiled for February 8th of 2010.**

Figure (9a, b, c) shows that the seismic activity and its corresponding electric one have drastically decreased therefore ellipses disappear and hyperbolas reappear as the main characteristic feature of the phase maps.

The phase maps which are presented in figures (6, 7, 8, 9a, b, c) are typical samples of the entire study time period. These serve the purpose to present, in a quick form, the phase map characteristic features change between a time period well before, very close and after the EQ occurrence time. Actually, for each day of the study period has been compiled the corresponding phase map. The presence of an ellipse is denoted by a blue bar at each day and compared to the time of occurrence of the EQs of Table – 1 (red bars) and the normalized raw and oscillating earth's electric field recorded from PYR and ATH monitoring sites. The latter is presented in the following figures (10, 11, 12, 13).

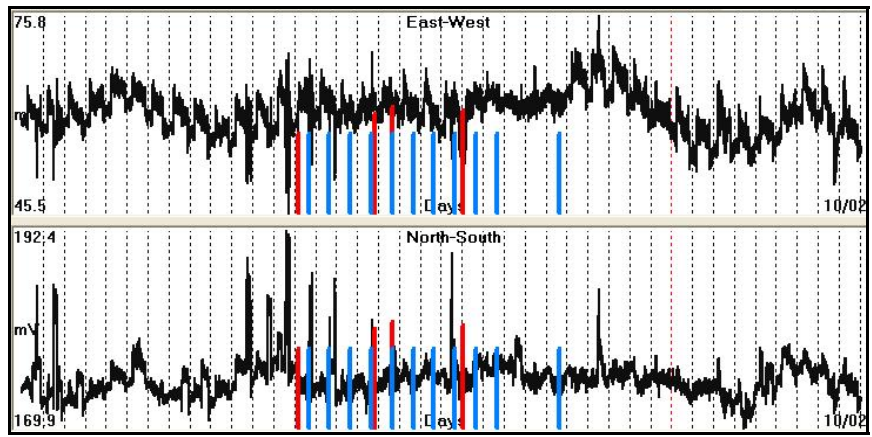

**Fig. 10. Daily presence of ellipses (blue bars) correlated to the EQ occurrence time (red bars) and to the normalized ATH raw data. Data spans from: January 1st 2010 to February 9th 2010.**

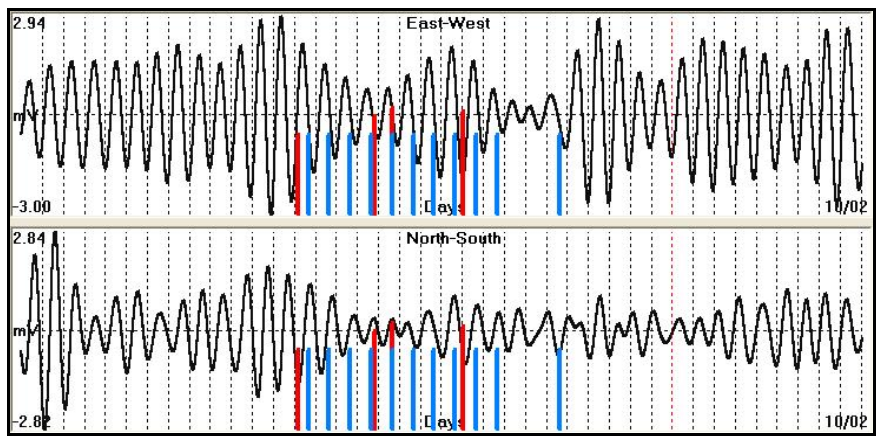

**Fig. 11. Daily presence of ellipses (blue bars) correlated to the EQ occurrence time (red bars) and to the normalized ATH oscillating data. Data spans from: January 1st 2010 to February 9th 2010.**

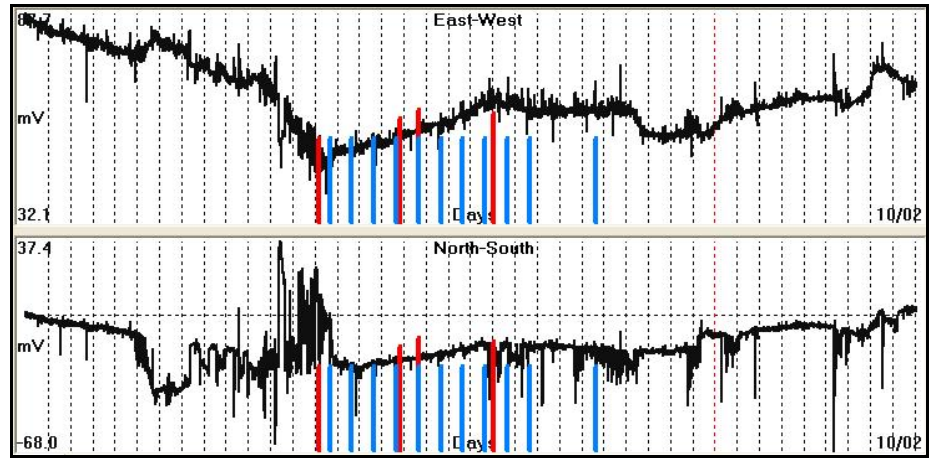

**Fig. 12.** Daily presence of ellipses (blue bars) correlated to the EQ occurrence time (red bars) and to the normalized PYR raw data. Data spans from: January 1st 2010 to February 9th 2010.

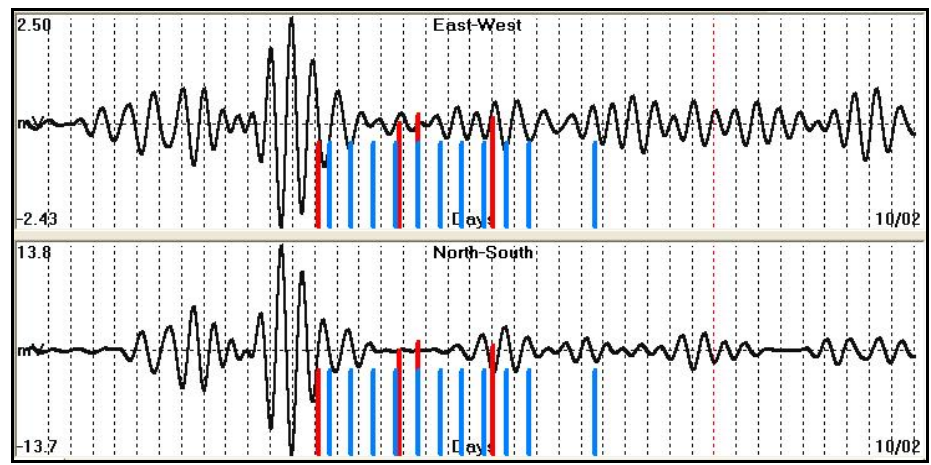

**Fig. 13.** Daily presence of ellipses (blue bars) correlated to the EQ occurrence time (red bars) and to the normalized PYR oscillating data. Data spans from: January 1st 2010 to February 9th 2010.

It is evident from figures (10, 11, 12, 13) that the elliptical feature of the phase maps is generated during the time period when an intense seismic activity is present in the regional Greek seimogenic area. However, the first seismic event (no. 1) has occurred at a depth of 88Km. Therefore, it is highly unlike to generate preseismic electric signals due to the fact that at these depths the lithosphere behaves more or less as a plastic medium than as a crystalline one. Consequently, the no. 1 seismic event has probably no contribution at all in the generation of the elliptical phase maps. Moreover its magnitude is only Ms = 5.0R and is the least one from all four of them.

The next seismic event (no. 2.) is located at a very long distance from the monitoring sites of PYR and ATH, quite larger compared to the ones of the seismic events no. (3) and no. (4). Moreover, this seismic event is still less in magnitude (Ms = 5.5R) compared to the magnitudes of no. (3) and no. (4). Therefore, by ignoring the seismic events of no. (1, 2) it is shown in figures (10, 11, 12, 13) that the presented characteristic ellipses in the compiled phase maps were generated four (4) days before the occurrence of the third seismic event (Ms = 5.7R). It is worth to notice that the ellipses (blue bars) continued to be present in the compiled maps even after the occurrence of the no. (3) seismic event. The latter practically suggested that the seismically activated area had not been stress discharged by the third seismic event. Therefore, it was highly probable that a next large seismic event could occur soon. This was verified by the occurrence of the forth seismic event of Ms = 5.6R at the very same epicentral area. After the occurrence of the last seismic event (no. 4) the ellipses disappear within a short time period.

In conclusion, the persistent presence of the ellipses, continuously for some days in the compiled phase maps, suggests the occurrence of a large EQ in the very near future. On the contrary, isolated ellipses (last right blue bar in figures 10, 11, 12, 13) indicate the occurrence of seismic events of minor magnitude (actually it coincides to an Ms = 4.5R seismic event located at SE of Greece, Karpathos Island, on January 26th, 2010).

## 2.2. Oscillating electric field of T = 14 days

A very similar behavior, as in paragraph (2.1) of the earth's oscillating electric field, has been observed prior to the occurrence time of large earthquakes at the T = 14 days component of the oscillating electric field too (Thanassoulas et al.

2009a). Therefore, the same analysis will be applied on the earth's electric field in order to test the validity of the method on the case of the Nafpactos EQs. In this specific case, a longer data set was analyzed since the corresponding earth's electric field wavelength is much larger than the one of the T = 1 day. The analyzed data set spans from December 10th 2009 to February 9th 2010. The normalized raw data and the corresponding oscillating component of the registered earth's electric field are presented in the following figures (14, 15, 16, 17) along with the EQs of Table – 1.

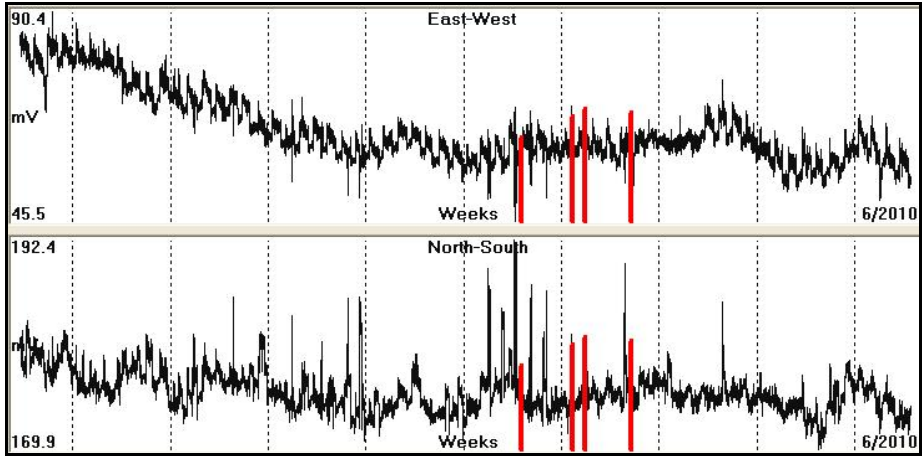

**Fig. 14.** ATH normalized raw data. Black line = normalized earth's electric field components. Red bars = occurred EQs of Ms ≥ 5.0R. Data spans from: December 10th 2009 to February 9th 2010.

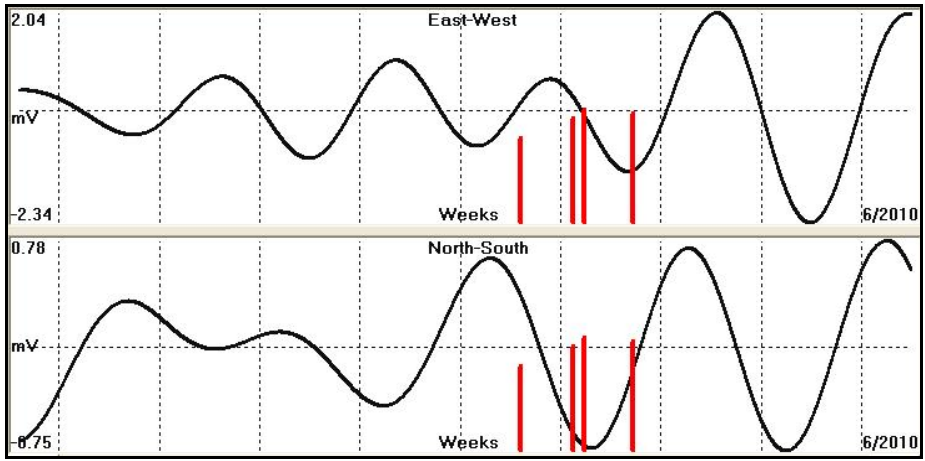

**Fig. 15.** ATH normalized oscillating data. Black line = normalized earth's electric field components. Red bars = occurred EQs of Ms ≥ 5.0R. Data spans from: December 10th 2009 to February 9th 2010.

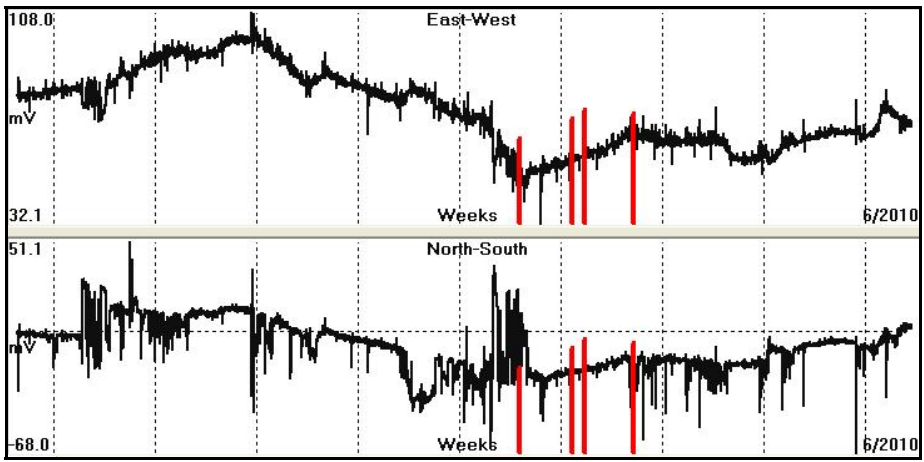

**Fig. 16.** PYR normalized raw data. Black line = normalized earth's electric field components. Red bars = occurred EQs of Ms ≥ 5.0R. Data spans from: December 10th 2009 to February 9th 2010.
8

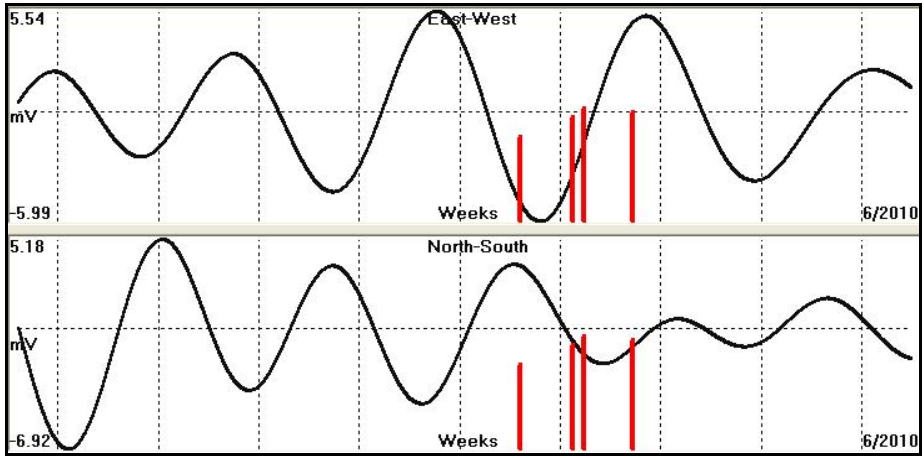

Fig. 17. PYR normalized oscillating data. Black line = normalized earth's electric field components. Red bars = occurred EQs of Ms ≥ 5.0R. Data spans from: December 10th 2009 to February 9th 2010.

From the data which are presented in figures (15) and (17) the following phase maps have been compiled and are presented in the figures (18, 19). Each map represents a phase map compiled for a period of time which starts 21 days backwards from the date indicated at the low left corner of each map. Moreover, a step of seven days has been used in the progressive compilation, so that there is an overlap of a wavelength (T = 14 days) for each successive map to its previous one.

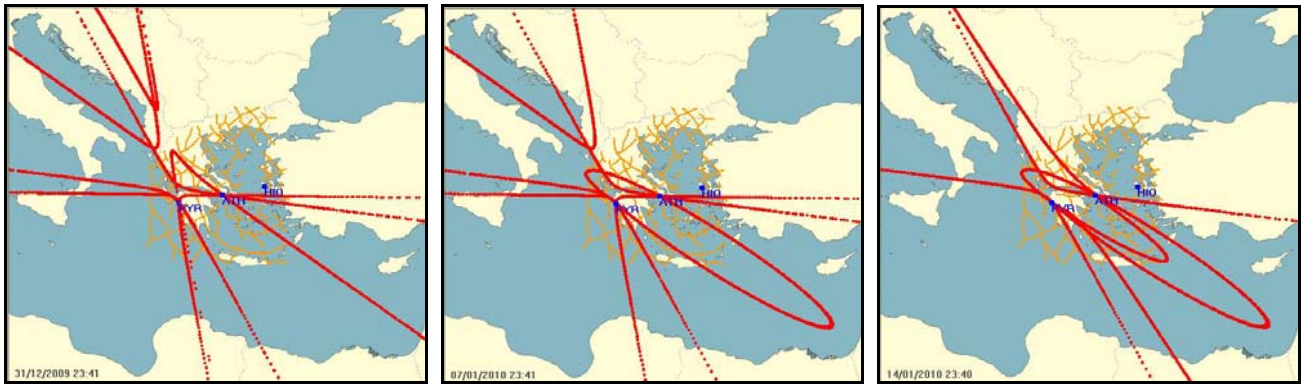

Fig. 18. Phase maps compiled for T = 14 days for the following dates: December 31st 2009 (left), January 7tht 2010 (middle), January 14th 2010 (right).

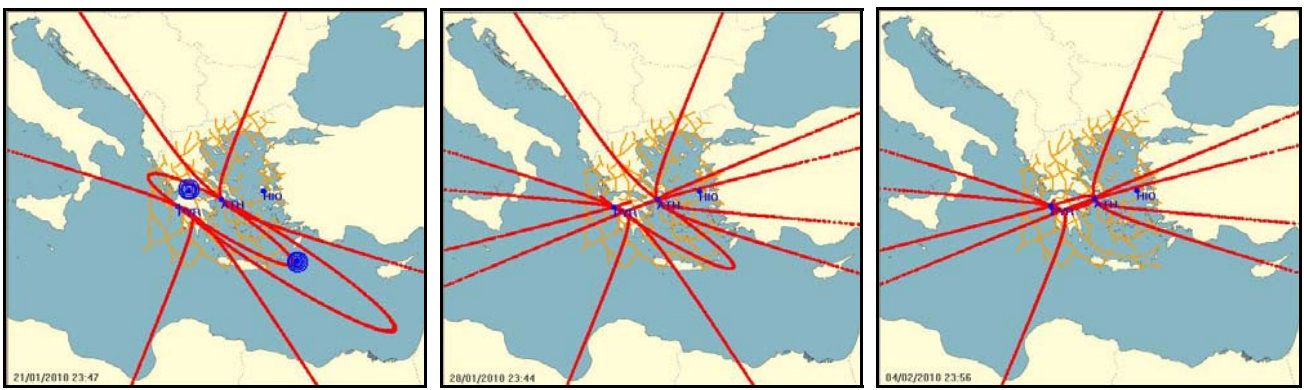

Fig. 19. Phase maps compiled for T = 14 days for the following dates: January 21st 2010 (left), January 28th 2010 (middle), February 4th 2010 (right).

It is clearly shown that the first phase map (fig. 18 left, 31/12/2009) suggests more or less normal stress load properties for the Greek regional seismogenic area due to the fact that only hyperbolas are observed. The second phase map (fig. 18 middle, 7/01/2010) indicates that the seismogenic area approaches a critical stress load level and therefore an ellipse is formed. The third phase map (fig. 18 right, 14/01/2010) indicates an increased stress load level (two ellipses are observed) and an EQ of Ms = 5.0R (no.1 of Table – 1) occurs on January 14th 2010. The forth phase map (fig. 19 left, 21/01/2010) still suggests that the seismogenic area is at a state of critical stress load conditions and therefore two larger EQs (Ms = 5.5R no. 2 and Ms = 5.7R no. 3 of Table – 1) take place. It must be pointed out that the forth EQ (Ms = 5.6R no. 4 of Table – 1) did occur on the January 22nd 2010 and therefore is not presented in this map. The fifth phase map (fig. 19 middle, 28/1/2010 suggests that the seismic stress load has been released. Therefore, ellipses disappear and hyperbolas again characterize the corresponding phase map. The same is observed in the sixth phase map (fig. 19 right, 4/2/2010).

## 3. Discussion - Conclusions

The aim of this work is to demonstrate the validity of the "strange attractor like" seismic electric precursor as a short-term earthquake prediction time of occurrence tool. To this end the specific methodology that is to compile phase maps of the earth's electric field monitored by two distant monitoring sites, was applied on Nafpactos EQs. The earth's electric field, recorded by PYR and ATH monitoring sites, long before, during and after the EQs occurrence, was used to compile the corresponding phase maps of the earth's electric field.

The specific methodology was applied on two distinct oscillating components of the earth's electric field. The first one is of T = 1 day and the second is of T = 14 days. It is understood that the frequency spectrum of the earth's electric field, which is generated by the activated physical mechanism in the focal area, is quite wide. Moreover, each separate frequency component of it will present the very same directional properties regarding the determined electric field intensity vector at each case. In other words its directional properties are independent from the corresponding frequency (Thanassoulas, 2007) provided that the lithosphere behaves as a homogeneous ground for the electric oscillating fields of very long wavelengths. Therefore, it is expected that since a seismogenic area has been stress charged close to critical level then the expected to be generated "strange attractor like" electric precursor will be present in all phase maps which are compiled for quite different (T) oscillating components.

That is one of the results of the present work. The corresponding phase maps were compiled for T = 1 day and T = 14 days and both of them presented similar behavior. Long before and after the EQs occurrence time the corresponding phase maps were characterized by hyperbolas while very close to the time of EQs occurrence the presence of ellipses was the main characteristic of the phase maps.

A question that may be raised now is: what the use of applying the methodology on different (T) electric field oscillating components is. To answer this just recall the basic optics physics law on illuminating physical objects by a light of specific wave length. The visible fine details of the physical object depend on the used light wavelength. The shorter the wavelength is the most of the object's fine details become visible. In the case of the "strange attractor like" electric seismic precursor, the shorter the wavelength of the earth's electric field is used the shorter the predictive window becomes. This is demonstrated in the following figures (20, 21).

The presence of an ellipse at a specific day within the study period has been assigned a value of (1) while its absence is assigned a value of (0). The time series of the ellipses presence (0s, 1s) has been weighted by a (1, 2, 1)*1.25 operator so that the ellipses presence time series are normalized up to a maximum value of (5). The latter is quite arbitrary and serves only the purpose of the presentation. The result of this operation is combined to the EQs occurrence time series. Figure (20) shows the correlation of the ellipses date of presence (strange attractor like) to the dates of the EQs occurrence.

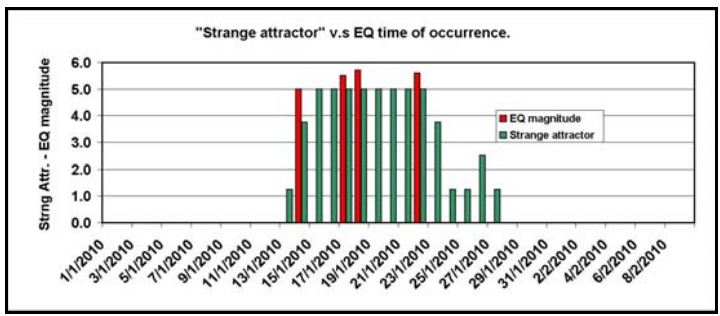

Fig. 20. Correlation of the "strange attractor presence date to the date of EQs occurrence from January 1st, 2010 to February 9th 2010.

If we neglect the first seismic event (no.1) which is assumed that does not contribute to the generation of the electric field due to its large occurrence depth, and the second one (no.2) due to its smaller magnitude and larger distance (compared to seismic events no. 3, 4) from the monitoring sites, then the following figure (21) is generated.

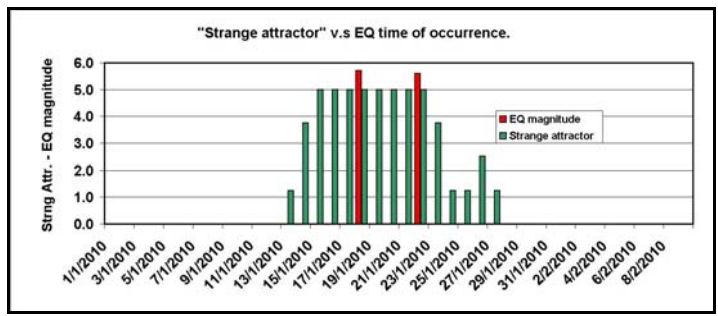

Fig. 21. Correlation of the "strange attractor presence to the date of EQs occurrence from January 1st, 2010 to February 9th 2010 after omitting seismic events no. (1) and no. (2).

What is shown in figure (21) that is very important, in terms of short-term (time) earthquake prediction, is the fact that the "strange attractor like" seismic precursor starts to show up only four days before the main event. The latter complies quite well with the previous presentations on the same topic by Thanassoulas et al. (2008a, b 2009a, b). Furthermore, after the occurrence of the main event (no. 3 of Table – 1) the strange attractor does not disappear as it should do, since the focal area theoretically is expected to have discharge the most of the stored strain energy, but still persists until a second similar in magnitude seismic event takes place(no. 4 of Table – 1). After the last seismic event the "strange attractor like" seismic electric precursor finally disappears.

For the case of the phase maps which have been compiled for the oscillating earth's electric field component of T = 14 days, the "strange attractor like" electric seismic precursor is initiated on the period of time from December 17th 2009 to January 7th 2010 (fig. 18, middle). Consequently, the time of occurrence of the pending EQ cannot be estimated with accuracy less than the used wavelength (in days). If we accept that the main seismic event will take place within a time period of a few wavelengths of the used (T), then it is obvious that the calculated time window for the expected EQ is extended much more than what was estimated for the case of T = 1 day.

Consequently, the use of progressively shorter periods (starting from large ones) of the oscillating earth's electric field for the compilation of the corresponding phase maps, suggests a scheme of progressively shortening of the estimated time window for the occurrence of a large EQ, being still in the frame of the short-term earthquake prediction.

The selection of the components of the used oscillating earth's electric field (T = 1, 14 days) are justified by the fact that are triggered by the corresponding lithospheric oscillation (and triggered large scale piezoelectricity) due to earth's tides. So, an extension of this work could be to investigate the presence of the "strange attractor like" electric seismic precursor for the case of the electric component of T = 12 hours, triggered by the corresponding earth's tide components of K2, M2, S2, and N2 in the range of T = 12 hours.

In conclusion, the "strange attractor like" seismic electric precursor was initiated four days before the Nafpactos EQs occurrence time. Not long before, nor long after the EQs occurrence, no similar seismic precursor was detected. That is a quite strong validation of the method. It indicated clearly that after the main event of no. (3), of Table – 1, a second EQ of similar magnitude was highly probable (the no. 4 of Table – 1). Its resolving capability for the time window expected, when a large EQ will occur, decreases as long as the wavelength of the used oscillating electric field used for its calculation increases.

Taking into account the results of this work along with the earlier similar results presented by Thanassoulas et al. (2008a, b, 2009a, b) it seems that the "strange attractor like" electric seismic precursor is a promising predictive tool in short-term earthquake prediction.